\documentclass
[%
aps,
reprint,
showpacs,
preprintnumbers,
amsmath,
amssymb,
pre,
]{revtex4-2}

\usepackage{graphicx}
\usepackage{dcolumn}
\usepackage{bm}
\usepackage{float}
\usepackage{subfigure}
\usepackage{color,soul}
\usepackage{lipsum}
\usepackage{lineno}
\usepackage{cancel}
\usepackage[normalem]{ulem}
\usepackage[section]{placeins}

\begin{document}
\preprint{}

\title{Effects of initial spin orientation on the generation of polarized electron beams from laser wakefield acceleration in plasma }

\author{L. R. Yin$^{1}$}
\author{X. F. Li$^{2}$}\email{xiaofengli@siom.ac.cn}%
\author{Y. J. Gu$^{3}$}
\author{N. Cao$^4$}
\author{Q. Kong$^{1}$}\email{qkong@fudan.edu.cn}%
\author{M. B\"uscher$^{5,6}$}
\author{S. M. Weng$^{7,8}$}
\author{M. Chen$^{7,8}$}
\author{Z. M. Sheng$^{7,8,9}$}

\affiliation{$^1$ Key Laboratory of Nuclear Physics and Ion-beam Application (MOE), Institute of Modern Physics, Department of Nuclear Science and Technology, Fudan University, Shanghai 200433, China}
\affiliation{$^2$State Key Laboratory of High Field Laser Physics, Shanghai Institute of Optics and Fine Mechanics, Chinese Academy of Sciences, Shanghai 201800, China}
\affiliation{$^3$SANKEN (Institute of Scientific and Industrial Research), Osaka University, Mihogaoka 8-1, Ibaraki, Osaka 567-0047, Japan}
\affiliation{$^4$Sichuan Research Institute, Shanghai Jiao Tong University, Sichuan 610200, China}
\affiliation{$^5$Peter Gr\"unberg Institut (PGI-6), Forschungszentrum J\"ulich, Wilhelm-Johnen-Str. 1, 52425 J\"ulich, Germany }
\affiliation{$^6$Institut f\"ur Laser- und Plasmaphysik, Heinrich-Heine-Universit\"at D\"usseldorf, Universit\"atsstr. 1, 40225 D\"usseldorf, Germany}
\affiliation{$^7$Key Laboratory for Laser Plasmas (MoE), School of Physics and Astronomy,Shanghai Jiao Tong University, Shanghai 200240, China}%
\affiliation{$^8$Collaborative Innovation Center of IFSA, Shanghai Jiao Tong University, Shanghai 200240, China}
\affiliation{$^9$Tsung-Dao Lee Institute, Shanghai Jiao Tong University, Shanghai 200240, China}

\date{\today}

\begin{abstract}
The effects of the initial spin orientation on the final electron beam polarization via laser wakefield acceleration in pre-polarized plasma are investigated theoretically and numerically. From a variation of the initial spin direction, the spin dynamics of the electron beam is found to depend on the self-injection mechanism. The effects of wakefields and laser fields are studied using test particle dynamics and particle-in-cell simulation based on the Thomas-Bargmann-Michel-Telegdi equation, respectively. Compared to the case of transverse injection, the scheme of longitudinal injection is more favorable to obtain a highly polarization electron beam.
\end{abstract}



\maketitle

\section{INTRODUCTION}
Spin-polarized electron beams have remarkable applications in nuclear and particle physics \cite{C.Glashausser1979,D.Androic2018,G.MoortgatPick2008,M.Burkardt2010,E.S.Ageev2005}, such as testing the standard model of particle physics \cite{D.Androic2018,G.MoortgatPick2008}, exploring the structure of subatomic particles \cite{M.Burkardt2010}, or detecting nuclear spin structures \cite{E.S.Ageev2005}. In conventional accelerators, such as electron storage rings, the Sokolov-Ternov effect requires approximately several hours for establishing polarization \cite{A.A.Sokolov1967,S.R.Mane2005}. On the other hand, the conventional electron accelerators are limited by the breakdown of radio-frequency cavities ($100$ MV/m) and are large-scale as well as expensive \cite{M.Chodorow1955,H.Wiedemann2015}. Owning to extremely high accelerating gradients above $100$ GV/m, laser-wakefield acceleration (LWFA) has attracted growing attention, where the acceleration process can be accomplished within a few picoseconds \cite{T.Tajima1979,E.Esarey2009}.

With the continuous improvement of theories and experiments on LWFA \cite{C.Geddes2004,X.Wang2013,J.Osterhoff2008,J.P.Palastro2008,E.Brunetti2010,W.P.Leemans2006,A.Gonsalves2019,A.Picksley2023,A.vonBoetticher2023,S.S.Baturin2023,R.Sandberg2024} as well as the development of pre-polarized plasma targets \cite{D.Sofikitis2017,D.Sofikitis2018}, generation of spin-polarized electron beams with high-quality via different injection mechanisms has received considerably interest \cite{M.Wen2019,Y.T.Wu2019,Y.T.Wu2019E,Y.T.Wu2020,Z.Nie2021,Z.Nie2022,T.Sun2022,Z.Gong2023,S.Bohlen2023}. Wen \emph{et al.} proposed that high-current polarized electron beams could be produced through the density-transition injection in a pre-polarized gas plasma generated by laser-induced photo-dissociation \cite{M.Wen2019}. Wu \emph{et al.} showed that the high spin-polarized electron beams could be achieved by using vortex Laguerre-Gaussian lasers \cite{Y.T.Wu2019,Y.T.Wu2019E,Y.T.Wu2020}. Nie \emph{et al.} proposed to generate electron beams with high polarization based on the ionization-induced injection mechanism \cite{Z.Nie2021,Z.Nie2022}. Recently, it was proposed that attosecond electron bunches with high spin-polarization could be obtained by using a radially polarized laser \cite{T.Sun2022}. Moreover, based on the colliding-pulse injection mechanism, quasi-monoenergetic polarized electron beams were expected to be produced with commercial $10$TW laser systems in a pre-polarized plasma \cite{Z.Gong2023,S.Bohlen2023}.

Among the above studies, the self-injection mechanism has been paid special attention since it is straightforward and can rather easily be implemented experimentally. When an ultra-intense laser pulse propagates into underdense plasma, electrons near the laser axis are pushed sidewards and form a blowout regime behind the laser. Some electrons can be accelerated after they are captured in the tail of bubble \cite{E.Esarey2009}. According to the different electron trajectories, the self-injection process can be divided as transverse injection and longitudinal injection \cite{S.Corde2013}. For the transversal case, the trapped electron initially located at the transverse radii of bubble regime \cite{Kostyukov2009} and its spin is dominantly affected by the bubble fields \cite{H.C.Fan2022}. For the longitudinal case, the captured electrons initially locate near the laser axis \cite{F.Y.Li2013} and their spins are mainly affected by the laser fields \cite{L.R.Yin2024}.

In the experimental scheme suggested in Ref. \cite{M.Buscher2020}, the pre-polarized plasma is formed by a dedicated laser beam and it is difficult to perfectly align with the accelerating laser. Therefore, the electron spin direction has an angle to the laser axis. In this paper, this phenomenon is investigated taking into account both the bubble fields and laser fields. The theoretical analysis is outlined in Sec. II. Numerical simulation results are presented in Sec. III, where transversal and longitudinal self-injections are discussed, respectively. The conclusions are given in Sec. IV.

\section{Theoretical Analysis}

The electron spin can be treated as a quasi-classical state with a vector $\boldsymbol{s}$ which has an absolute value of $1$. If an electron with velocity $\boldsymbol{v}$ moves in an electromagnetic field, the precession of its spin vector $\boldsymbol{s}$ can be calculated through the Thomas-Bargmann-Michel-Telegdi (TBMT) equation \cite{L.H.Thomas1926,S.R.Mane2005}
\begin{equation}\label{eq1}
\mathrm{d}\boldsymbol{s}/\mathrm{d}t=\boldsymbol{\Omega}\times\boldsymbol{s},
\end{equation}
with the spin precession frequency
\begin{equation}\label{eq2}
\begin{split}
\boldsymbol{\Omega}=\frac{e}{m_{e}}\Bigg[&\left(a_{e}+\frac{1}{\gamma}\right)\boldsymbol{B}-\frac{a_{e}\gamma}{\gamma+1}\boldsymbol{v}\cdot \boldsymbol{B}\frac{\boldsymbol{v}}{c^{2}}
\\&-\left(a_{e}+\frac{1}{\gamma+1}\right)\frac{\boldsymbol{v}}{c^{2}}\times\boldsymbol{E}\Bigg].
\end{split}
\end{equation}
Here, $m_{e}$, $e$, and $\gamma$ are the mass, charge, and Lorentz factor of the electron, $a_{e}\approx1.16\times10^{-3}$ is the dimensionless anomalous magnetic moment, $c$ is the speed of light in vacuum, $\boldsymbol{B}$ is the magnetic field, and $\boldsymbol{E}$ is the electric field in the laboratory frame. During LWFA, the effects of the radiation reaction, Stern-Gerlach and Sokolov-Ternov can be ignored according to Ref. \cite{J.Thomas2020}.

\begin{figure}[t]
	\includegraphics[width=0.48\textwidth]{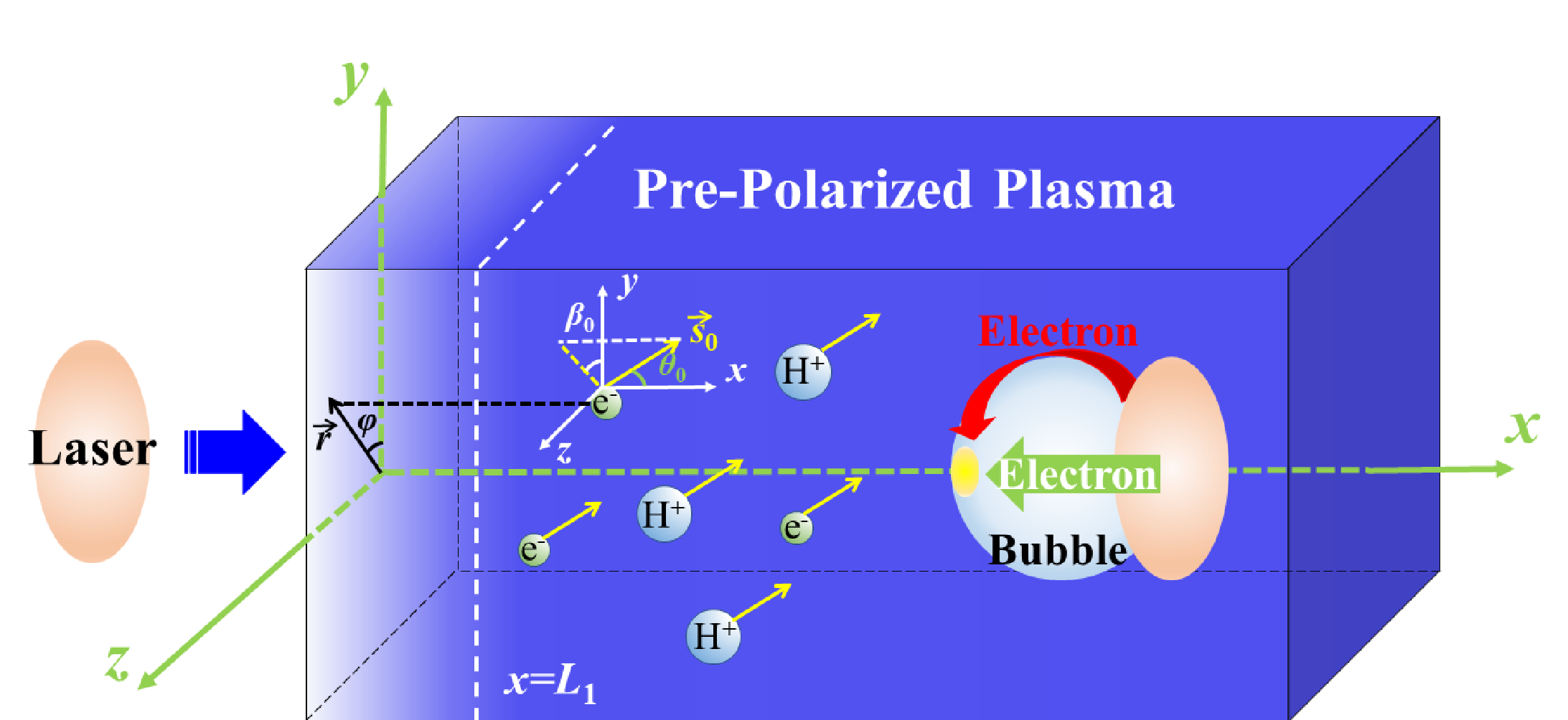}
	\caption{Schematic of the interaction of laser and pre-polarized plasma. The longitudinal profile of the plasma density is consist of a transition from $0$ to $n_{0}$ with length $L_{1}$ and a plateau with $n_{0}$. The laser is focused at the left-hand boundary of the plasma. The initial spin status is defined by $ \theta_{0}$ and $\beta_{0}$, which is denoted by the yellow arrows. The self-injection mechanism for accelerating electrons can be divided into the transverse (red arrow) and longitudinal scheme (green arrow) based on their trajectories.}\label{fig1}
\end{figure}

For simplicity, the initial degree of polarization in the plasma is assumed to be $100\%$, as the effect of the initial spin direction is the main focus of this study. As presented in Fig. \ref{fig1}, a laser with linear polarization along the $y$-direction propagates in the plasma along the $x$-direction. For a general case, the electron spin direction has angle with the $x$-direction, which is denoted by $\theta_{0}$ and $\beta_{0}$ in Fig. \ref{fig1}. Then the initial spin of electrons $\boldsymbol{s}_{0}$ can be written as $\boldsymbol{s}_{0}=\cos\theta_{0}\bm{\mathrm{\hat{i}}}+\cos\beta_{0}\sin\theta_{0}\bm{\mathrm{\hat{j}}}+\sin\beta_{0}\sin\theta_{0}\bm{\mathrm{\hat{k}}}$, as indicated by the yellow arrows in Fig. \ref{fig1}.

When electrons are captured and accelerated in the blowout regime, their depolarization mainly happens in the injection process rather than the acceleration process. In the case of transverse injection, the precession of electron spin is mainly divided into two stages for the injection process, according to Ref. \cite{H.C.Fan2022}. In the first stage, the electrons arrive at the bubble sheath in transverse direction, and their spins are mainly affected by the magnetic field $B_{\varphi}$ and electric field $E_{r}$. In the second stage, the electrons are located at the tail of the bubble and their spins are mainly influenced by the electric field $E_{x}$. This results in a spin precession frequency about electrons $\boldsymbol{\Omega}\approx \boldsymbol{\Omega_{\varphi}}$ according to Eq. (\ref{eq1}) with constant direction of $\boldsymbol{\Omega}$ during the injection process. Here, cylindrical coordinates are used because the distribution of electromagnetic field is approximately axially symmetry in the bubble regime.

\begin{figure}[t]
\includegraphics[width=0.46\textwidth]{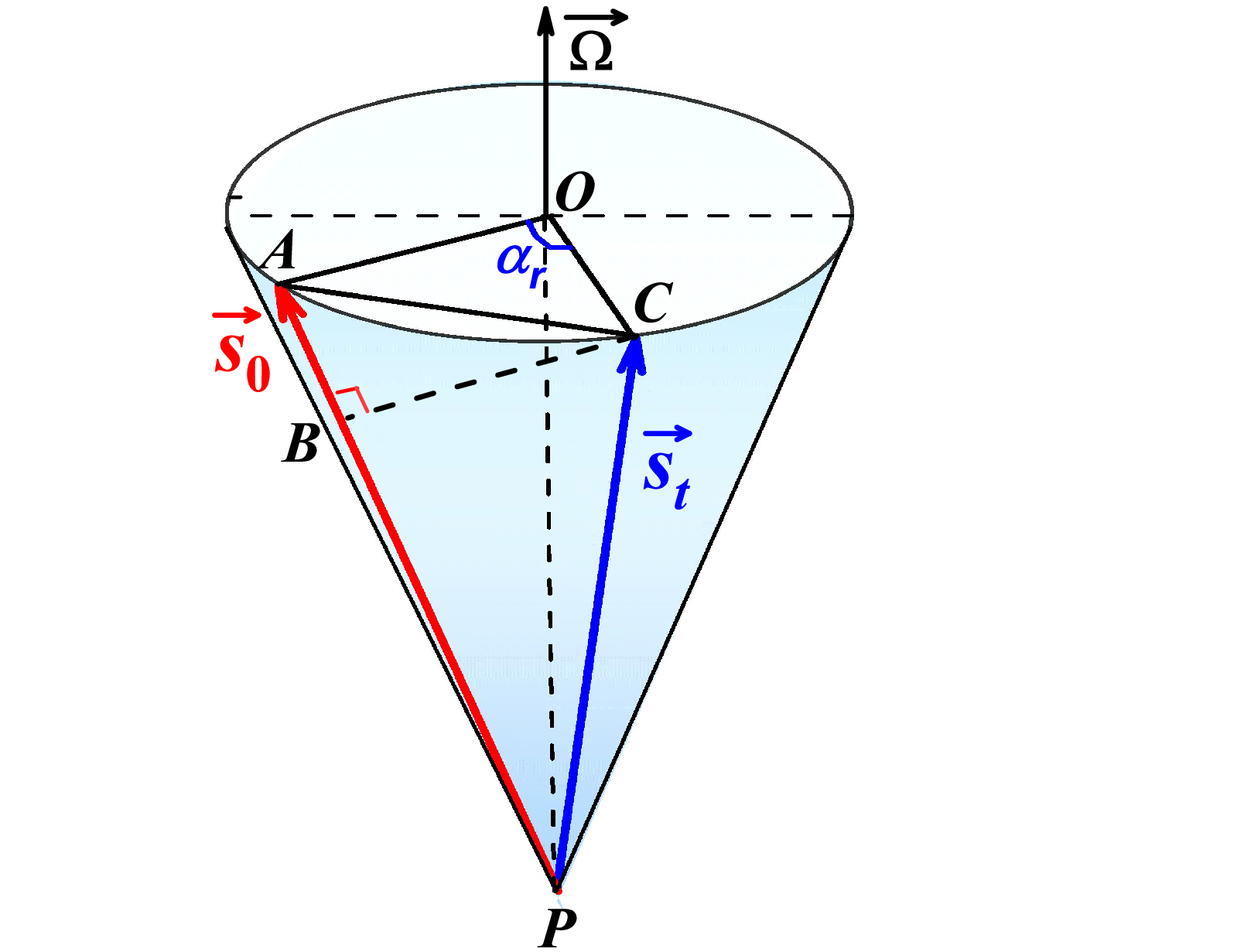}
\caption{Schematic diagram of electron spin evolution with a fixed direction of precession frequency $\boldsymbol{\Omega}$. The electron spin at initial ($\boldsymbol{s_0}$) and final time ($\boldsymbol{s_t}$) are denoted by the red and blue arrows, respectively. $\alpha_{r}$ represents the rotation angle during $\bigtriangleup t$ in the $AOC$ plane.}\label{fig2}
\end{figure}

According to Eq. (\ref{eq1}), the precession of electron spin is treated as the rotational motion of the spin vector $\boldsymbol{s}$ around the direction of precession frequency $\boldsymbol{\Omega}$. Since the direction of $\boldsymbol{\Omega}$ does not change with time, the spin rotation depends on the strength of $\boldsymbol{\Omega}$ and the rotational motion defines the plane $AOC$, as displayed in Fig. \ref{fig2}. The red and blue arrows represent the initial spin vector $\boldsymbol{s_{0}}=s_{0\psi}\boldsymbol{\hat{\psi}}+s_{0\kappa}\boldsymbol{\hat{\kappa}}$ and final spin vector $\boldsymbol{s_{t}}=s_{t\psi}\boldsymbol{\hat{\psi}}+s_{t\kappa}\boldsymbol{\hat{\kappa}}$, where $\boldsymbol{\hat{\psi}}$ and $\boldsymbol{\hat{\kappa}}$ are unit vectors parallel and perpendicular to the direction of $\boldsymbol{\Omega}$, respectively. The rotation angle in the rotation plane between the initial status (PA) and final time (PC) is defined as $\alpha_{r}$. Therefore, the component of the final spin $\boldsymbol{s_{t}}$ parallel to the direction of initial spin $\boldsymbol{s_{0}}$ for an electron can be written as,
\begin{equation}\label{eq3}
s_{\parallel}=\boldsymbol{s_{t}}\cdot\boldsymbol{\hat{s}_{0}}
=1-2|s_{0\kappa}\boldsymbol{\hat{\kappa}}|^{2}\sin^{2}\left(\frac{\alpha_{r}}{2}\right).
\end{equation}

For the accelerated electrons in the bubble field, considering $\boldsymbol{\Omega}\approx \boldsymbol{\Omega_{\varphi}}$, the component of the initial spin vector $\boldsymbol{s_{0}}$ parallel to the direction of $\boldsymbol{\hat{\psi}}$ can be written as
\begin{equation}\label{eq4}
\begin{split}
|s_{0\psi}\boldsymbol{\hat{\psi}}|&=|\boldsymbol{s_{0}}\cdot \boldsymbol{\hat{\varphi}}|\\&=|(\cos\theta_{0}\hat{\boldsymbol{i}}+\cos\beta_{0}\sin\theta_{0}\hat{\boldsymbol{j}}+\sin\beta_{0}\sin\theta_{0}\hat{\boldsymbol{k}})\cdot\boldsymbol{\hat{\varphi}}|
\\&=|-\sin\theta_{0}\cos\beta_{0}\sin\varphi+\sin\theta_{0}\sin\beta_{0}\cos\varphi|. \end{split}
\end{equation}
Considering $|s_{0\kappa}|=\sqrt{1-|s_{0\psi}|^{2}}$, Eq. (\ref{eq3}) can be rewritten as,
\begin{equation}\label{eq5}
\begin{split}
s_{\parallel}&=\cos\alpha_{r}+\\&2\sin^{2}\left(\frac{\alpha_{r}}{2}\right)\sin^{2}\theta_{0}(\sin\beta_{0}\cos\varphi-\cos\beta_{0}\sin\varphi)^{2}.
\end{split}
\end{equation}
It can be seen that $s_{\parallel}$ is related to the initial polarization direction, \emph{i.e.}, pre-polarization angle $\theta_{0}$ and azimuth angle $\beta_{0}$, and $s_{\parallel}$ also depends on the electrons position and electromagnetic field, as presented in $\varphi$ and $\alpha_r$.

The polarization of the accelerated beam can be derived as the average value of the injected electron spins. For the case of transverse injection scheme, the electrons are initially located in the region around the transverse radius, from $R_{0min}$ to $R_{0max}$. The electrons at the same distance from the laser axis $r$ are injected into the bubble at the same time. According to Eq. (\ref{eq5}), the polarization of the injected electrons can be derived as
\begin{equation}\label{eq6}
\begin{split}
\langle s_{\parallel}\rangle&=\frac{\int_{R_{0min}}^{R_{0max}}\int_{-\pi}^{\pi}{s_{\parallel}}\cdot r\mathrm{d}\varphi\mathrm{d}r}{\int_{R_{0min}}^{R_{0max}}\int_{-\pi}^{\pi} r\mathrm{d}\varphi\mathrm{d}r}\\&=\frac{1}{2}\left[\langle\cos\alpha_{r}\rangle\left(2-\sin^{2}\theta_{0}\right)+\sin^{2}\theta_{0}\right],
\end{split}
\end{equation}
where $\langle\cos\alpha_{r}\rangle=\int_{R_{0min}}^{R_{0max}}\cos\alpha_{r}\cdot r\mathrm{d}r/(R_{0max}^{2}-R_{0min}^{2})$. Noteworthy, the value of $\langle s_{\parallel}\rangle$ is related to $\theta_{0}$. However, it is independent of $\beta_{0}$, which can be always set to zero based on a transformation of coordinates. Besides,  for the case with $\theta_{0}=0^{\circ}$, $\langle s_{\parallel}\rangle=\cos\alpha_{r}\in[-1,1]$, while $\langle s_{\parallel}\rangle=1/2(1+\cos\alpha_{r})\in[0,1]$ in the case with $\theta_{0}=90^{\circ}$. More importantly, the electron polarization $\langle s_{\parallel}\rangle$ increases monotonically with $\theta_{0}$ in the range of $[0,\pi/2]$, which indicates that the initial status of electrons spin can affect the final polarization.

\section{Simulation methods and discussion}

In order to confirm the above theoretical analysis, a series of test particle dynamics and Particle-In-Cell (PIC) simulation are used to study the spin motion of electrons. The PIC simulations are performed with a modified version of the EPOCH code \cite{T.D.Arber2015}, which includes the spin evolution module based on the TBMT equation via the Boris pusher method \cite{X.F.Li2021}. The laser propagates in the $x$-direction with linear polarization along the $y$-direction and a transversal Gaussian envelope
\begin{equation}\label{eq7}	E_y=\frac{E_{0}w_{0}}{w\left(x\right)}\mathrm{exp}\left[-\frac{y^{2}+z^{2}}{w\left(x\right)^{2}}-\frac{\left(t-\tau\right)^{2}}{\left(0.5\tau\right)^{2}}\right]\!\cos\left(\varphi\right),
\end{equation}
where the laser wavelength is $\lambda=800$ $\mathrm{nm}$, $w(x)=w_{0}\left[1+\left(x-x_{0}\right)^{2}/z_{R}^{2}\right]^{1/2}$, $z_{R}=\pi w_{0}^{2}/\lambda$, $\tau$ is the laser pulse duration and the normalized laser amplitude is $a_{0}=eE_{0}/m_{e}\omega c$. The laser pulse is focused at the left edge of the plasma target ($x_{0}=30\lambda$). The initial longitudinal profile of the pre-polarized plasma is an up-ramp followed by a plateau \cite{F.Y.Li2013}, as presented in Fig. \ref{fig1}. The normalized laser amplitude $a_{0}$, the spot size is $w_{0}$, the length of up-ramp transition $L_{1}$ and plasma density $n_{0}$ are different in the case of transverse and longitudinal injection. The $2$D simulation box is $200\lambda(x)\times120\lambda(y)$ with resolution $\mathrm{d}x=0.02\lambda$ and $\mathrm{d}y=0.08\lambda$. A moving window is used to accelerate the electrons sufficiently, and there are $4$ pseudo-particles per cell for each particle species.

\subsection{Transverse self-injection mechanism}

In the case of the transverse self-injection mechanism, the accelerated electrons are initially located at the transverse radii and the injection process is affected by the distribution of the electromagnetic field, which is impacted by the bubble geometry. Based on the work of Li \emph{et al.} \cite{X.F.Li2015}, the electromagnetic field of an ellipsoidal bubble can be written as
\begin{subequations}\label{8}
\begin{equation}\label{8a}
	E_x=\frac{\eta^2}{\eta^2(1-v_\mathrm{b}^2)+2}\xi,   \vspace{1ex}
\end{equation}
\begin{equation}\label{8b}
	E_y=\frac{2-\eta^2 v_\mathrm{b}^2}{2 \eta^2(1-v_\mathrm{b}^2)+4}y,
\end{equation}
\begin{equation}\label{8c}
	E_z=\frac{2-\eta^2 v_\mathrm{b}^2}{2\eta^2(1-v_\mathrm{b}^2)+4}z,
\end{equation}
\begin{equation}\label{8d}
	B_x=0,
\end{equation}
\begin{equation}\label{8e}
	B_y=\frac{v_\mathrm{b} \eta^2}{2 \eta^2(1-v_\mathrm{b}^2)+4}z,
\end{equation}
\begin{equation}\label{8f}
	B_z=-\frac{v_\mathrm{b} \eta^2}{2 \eta^2(1-v_\mathrm{b}^2)+4}y,
\end{equation}
\end{subequations}
where $\xi=x-v_{\mathrm{b}}t$, $v_{\mathrm{b}}=\sqrt{1-\gamma_{\mathrm{b}}^{-2}}$ is the bubble phase velocity and $\gamma_{\mathrm{b}}=0.45\sqrt{n_{c}/n_{0}}$\cite{I.Kostyukov2004,C.B.Schroeder2011}. To confine the field distribution inside the bubble, a modified factor $f(r)=\left[\tanh({R_{\parallel}/d-r/d})+1\right]/2$ is used, $r=\sqrt{\xi^2+(y^2+z^2)/\eta^2)}$ and $d$ is the width of the electron sheath. In this work, $d=0.5$ is used. The aspect ratio $\eta=R_{\perp}/R_{\parallel}$ is defined to describe the geometry of the bubble, where $R_{\parallel}$ and $R_{\perp}$ are the longitudinal and transverse radii, respectively. They are calculated directly through $2$D PIC simulation. Following this definition, $\eta<1$, $\eta=1$ and $\eta>1$ indicate a prolate spheroid, a sphere and an oblate spheroid, respectively.

The electron dynamics in the bubble field is calculated by the fourth-order Runge-Kutta method based on the relativistic Newton-Lorentz equation $\mathrm{d}\boldsymbol{P}/\mathrm{d}t=-e[\boldsymbol{E}+(\boldsymbol{P}/\gamma)\times \boldsymbol{B}]$, where $\gamma=1/\sqrt{1-v^{2}/c^{2}}$ is the relativistic factor of the electrons. Meanwhile, the spin precession of an electron is calculated according to the TBMT equation with the Boris-rotation method. The electrons are initially located at the front of the bubble, denoted as $(x_{0},y_{0},z_{0})$. The initial direction of electron spin $\boldsymbol{s}_0$ changes in the $xy$ plane in order to study the effect of initial spin on final polarization.

\begin{figure}[b]
\includegraphics[width=0.38\textwidth]{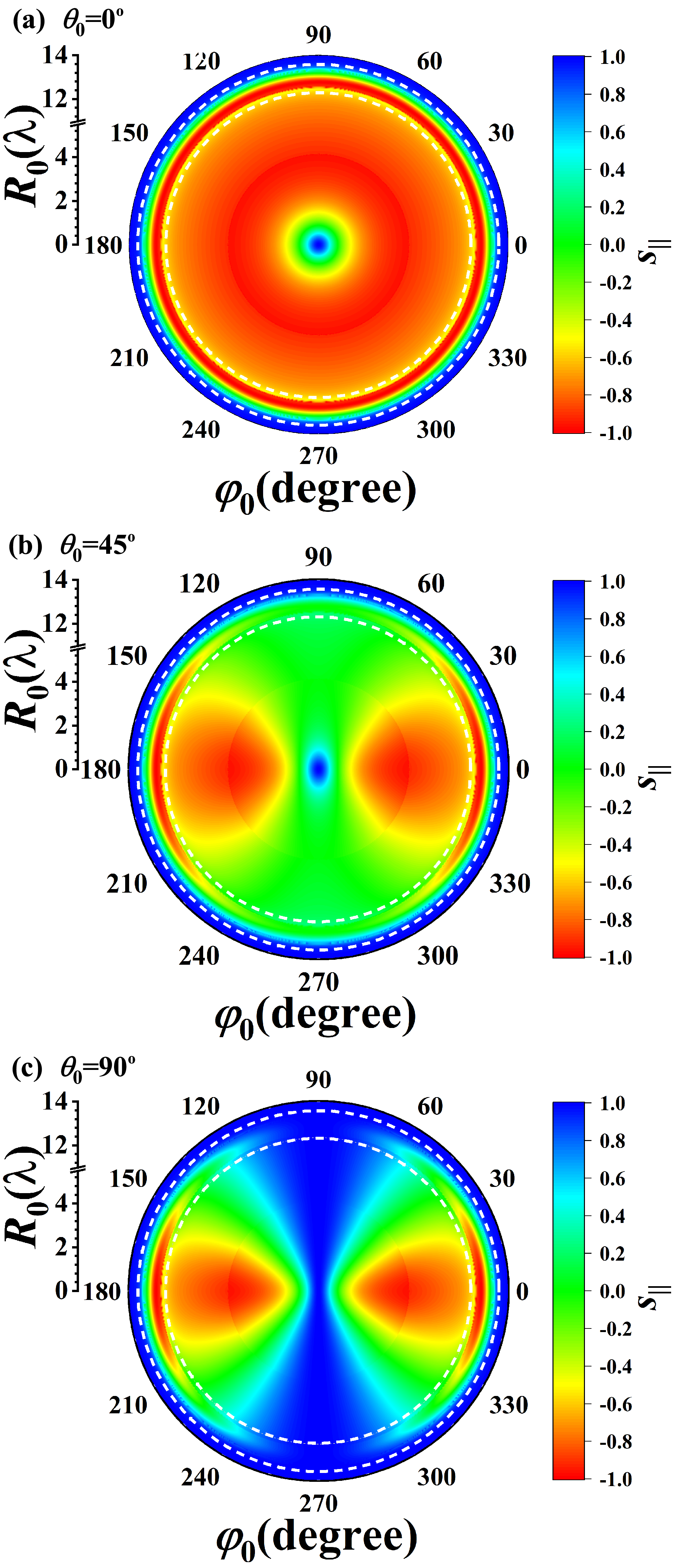}
\caption{Distributions of $s_{\parallel}$ in dependence on the initial electron position ($R_0, \varphi_0$) for different initial polarization directions $\theta_{0}=0^{^{\circ}}$(a), $45^{^{\circ}}$(b), and $90^{^{\circ}}$(c), respectively. Here, $R_0=\sqrt{y_0^2+z_0^2}$,  $\varphi_{0}=\tan^{-1}({z_{0}}/{y_{0}})$, and $\theta_{0}=\tan^{-1}({s_{0y}}/{s_{0x}})$. The laser parameters are $a_{0}=20$, $w_{0}=10\lambda$ and $\tau=21$fs and the plasma density is  $n_{0}=0.021n_c$. The bubble parameters are obtained through the PIC simulations, where $R_{\parallel}=10.31\lambda$, $R_{\perp}=12.34\lambda$ and $\eta=R_{\perp}/R_{\parallel}=1.20$. The electrons, initially located in the region between the white dashed lines from $R_{0min}$ to $R_{0max}$, can be captured by the bubble and obtain the high energy finally.  }\label{fig3}
\end{figure}

The distribution of electron spin $s_{\parallel}$ at final time with initial transverse position ($R_{0}, \varphi_{0}$) are presented in Fig. \ref{fig3}, where $R_0=\sqrt{y_0^2+z_0^2}$ and $\varphi_{0}=\tan^{-1}({z_{0}}/{y_{0}})$. $\theta_{0}=\tan^{-1}({s_{0y}}/{s_{0x}})$ denotes the initial spin direction. $\theta_{0}=0^{\circ}$ indicates that the initial spin is align with the laser-propagation direction. Electrons with final kinetic energy larger than $20$MeV are selected as the accelerated electrons. It turns out that the accelerated electrons are located around the transverse radius initially, as indicated by the white dashed circles in Fig. \ref{fig3}(a). More importantly, the distribution of $s_{\parallel}$ is axial symmetry. With changing the spin initial status $\theta_{0}$, the final distribution of $s_{\parallel}$ occurs obviously difference not only for acceleration electrons, as revealed in Figs. \ref{fig3}(b) and \ref{fig3}(c). The distribution of electron spin $s_{\parallel}$ does not change obviously for the electrons located at $z$-axis initially, while its value increases from $\theta_0=0^{\circ}$ to $\theta_0=90^{\circ}$ for the electrons located at $y$-axis initially. Then, it results in the polarization of the accelerating electron beam relying on the initial spin status.

According to the theoretical analysis in Sec. II, the spin procession of the accelerated electrons is determined by $\Omega_{\varphi}$, which leads to the rotation of electron spin around the direction of $\boldsymbol{\hat{\varphi}}$ and it induces that the distribution of $s_\parallel$ is isotropic at $\theta_{0}=0^{\circ}$ and is anisotropic at $\theta_{0}=90^{\circ}$ and $45^{\circ}$. Furthermore, $\langle s_\parallel\rangle$ is independent of the pre-polarization angle $\theta_{0}$ for the electrons located at the $z$-axis, while $\langle s_\parallel\rangle$ relies on $\theta_{0}$ for the electrons located at $y$-axis.

\begin{figure}[b]
	\includegraphics[width=0.40\textwidth]{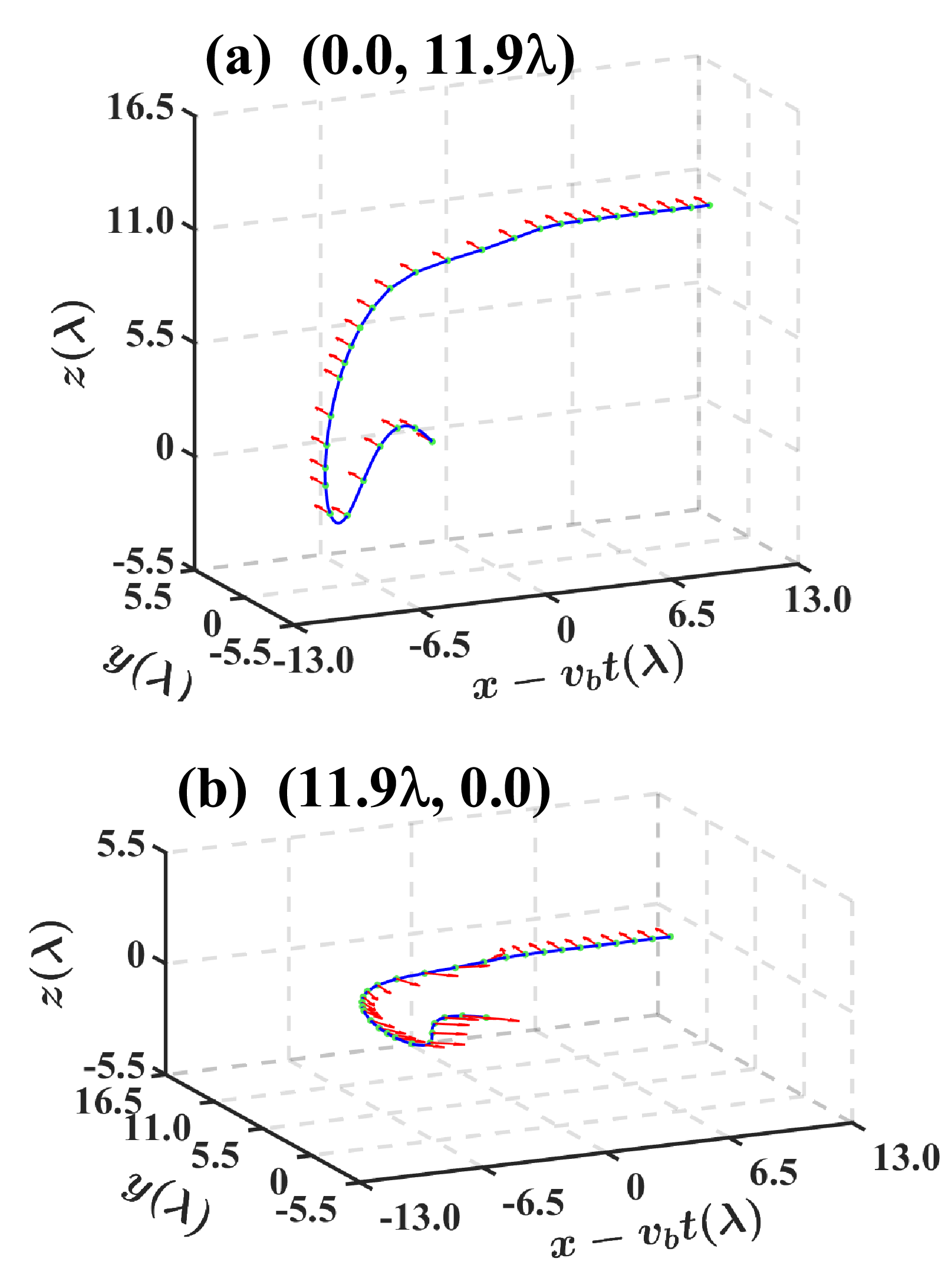}
	\caption{The history of electron trajectory (blue solid lines) and spin precession (red arrows) in the case of transverse injection. The initial positions ($y_0,z_0$) of the electron are at (0.0,$11.9\lambda$) (a) and ($11.9\lambda$,0.0)(b), respectively. The other parameters are same as those in Fig. 3(c).}\label{fig4}
\end{figure}

The electron trajectory explains this phenomenon thoroughly. The dynamics of trajectory and spin about two typical electrons with same distance from $x$-axis are presented in Figs. \ref{fig4}(a) and \ref{fig4}(b). The direction of electron spin located at the $y$-axis initially, \emph{i.e.,} $\theta_{0}=90^{\circ}$ and the other parameters are same with them in Fig. \ref{fig3}(c). As shown in Fig. \ref{fig4}(a), the spin of the accelerated electron, located at $z$-axis initially, does not fluctuate with time, because its spin direction is always aligned with the direction of precession frequency $\boldsymbol{\Omega}$. On the contrary, as revealed in Fig. \ref{fig4}(b), the accelerated electron is located at $y$-axis initially and its spin direction starts to rotate around the $z$-axis. It is revealed that the electron spin are mainly influenced by $\Omega_{z}$, which is consistent with the theoretical analysis.

\begin{figure}[t]
	\includegraphics[width=0.49\textwidth]{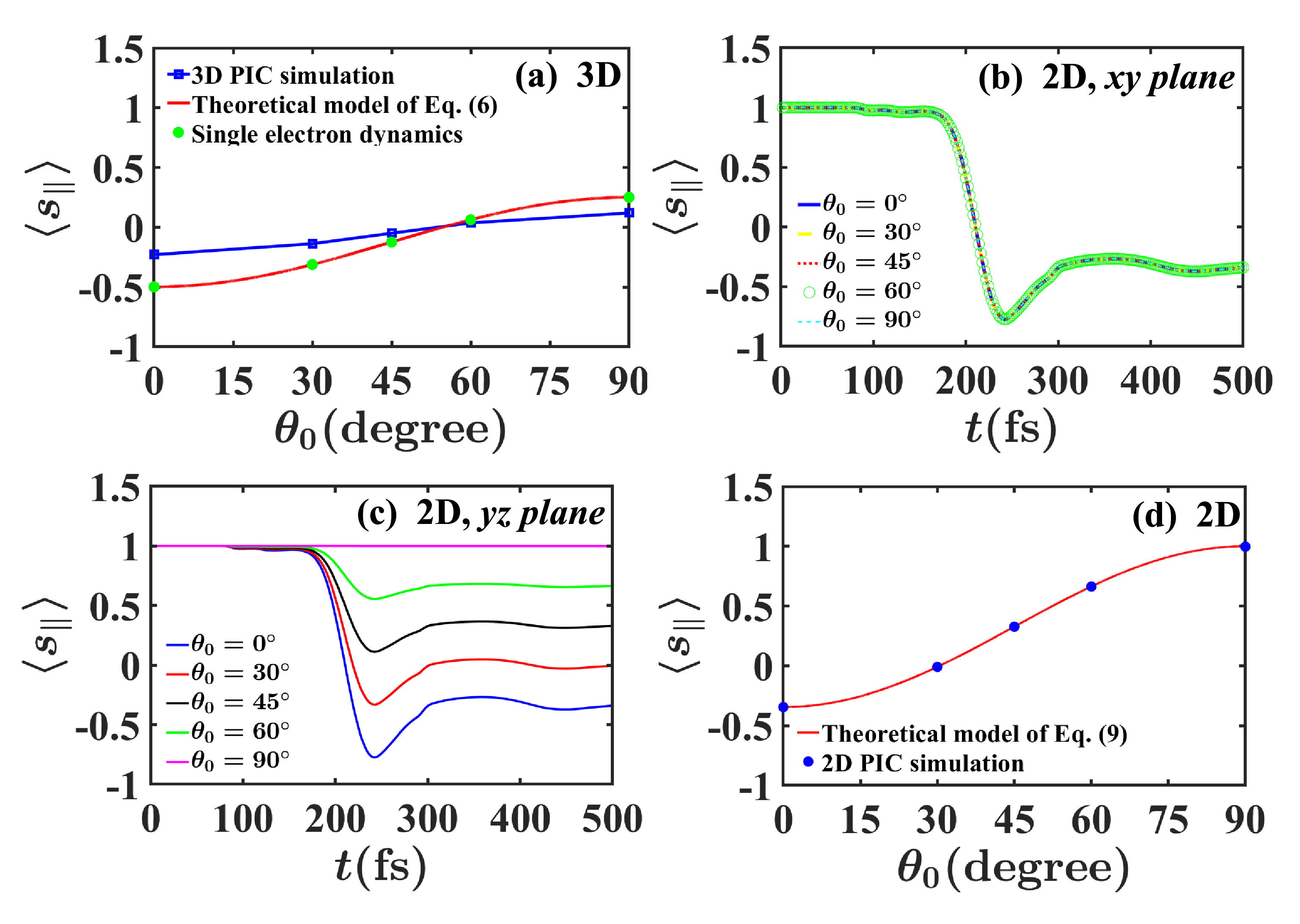}
	\caption{ (a) The relationship between the polarization $\langle s_{\parallel}\rangle$ of the accelerated electron beam and the pre-polarization angle $\theta_{0}$, calculated through $3$D PIC simulations, theoretical model (Eq. (6)) and single particle dynamics. The history of $\langle s_{\parallel}\rangle$ about the electron beam with different pre-polarization angle $\theta_{0}$ in 2D PIC simulation on $xy$ and $xz$ planes are displayed in (b) and (c), respectively. (d) The distribution of $\langle s_{\parallel}\rangle$ with $\theta_{0}$ for the accelerating electrons, calculated by the $2$D PIC simulation on $xz$ plane and the theoretical model (Eq. (9)). The laser parameters are $a_{0}=20$, $w_{0}=10\lambda$ and $\tau=21$ fs, respectively. The plasma density is $n_{0}=0.021n_{c}$ and $L_{1}=10\lambda$.}\label{fig5}
\end{figure}

The net polarization of a particle beam is defined as $P=\sqrt{ \langle s_x\rangle^2+\langle s_y\rangle^2+\langle s_z\rangle^2}$, where $\langle s_i\rangle$ is the average value in each direction. Figure \ref{fig5}(a) reveals the comparison for $\langle s_{\parallel}\rangle$ of the accelerated electrons between the $3$D PIC simulations, the theoretical prediction results based on Eq. (\ref{eq6}), and the results of test electron dynamics. The $\langle s_{\parallel}\rangle$ of the electron beam increases with increasing $\theta_{0}$. The results of test particle are fit with the theoretical analysis. Here, $\langle\cos\alpha_{r}\rangle$ is chosen based on the results of test particle dynamics. However, there is a tiny difference between theory and $3$D PIC simulations, which may result from the fact that the injection cross-section is not an ideal circular ring, owning to the bubble evolution.

For $2$D PIC simulation, Figs. \ref{fig5}(b) and \ref{fig5}(c) present the effect of $\theta_{0}$ on the polarization of accelerated electron beam $\langle s_{\parallel}\rangle$ in the $xy$ ($\varphi=0^{\circ}$ or $180^{\circ}$) and $xz$ planes ($\varphi={\pm}90^{\circ}$), respectively. When the simulation box is located in $xy$ plane (Fig. \ref{fig5}(b)), the history of $\langle s_\parallel \rangle $ is independent of initial spin status, because the direction of the spin precession is always perpendicular to the electron spin. According to Eq. (\ref{eq5}) and assuming $\beta_0=0^{\circ}$, the electron spin can be written as $s_{\parallel}=\cos\alpha_{r}$, which means that the electron spin is independent of the initial spin direction.

When the simulation box is located in the $xz$ plane and the polarization dynamics is affected by the initial status, as revealed in Fig. \ref{fig5}(c). Based on Eq. (\ref{eq5}) and assuming $\beta_{0}=0^{\circ}$, the electron spin can be written as $s_{\parallel}=\cos\alpha_{r}+2\sin^{2}(\alpha_{r}/2)\sin^{2}\theta_{0}$. Then, the polarization of the accelerated electrons is obtained as,
\begin{equation}
\begin{split}
\langle{s_{\parallel}}\rangle&=\frac{\int_{z_{0min}}^{z_{0max}}s_{\parallel}\cdot\mathrm{d}z}{\int_{z_{0min}}^{z_{0max}}\mathrm{d}z}\\&=\langle\cos\alpha_{r}\rangle\left(1-\sin^{2}\theta_{0}\right)+\sin^{2}\theta_{0},
\end{split}
\end{equation}
where $\langle\cos\alpha_{r}\rangle=\int_{z_{0min}}^{z_{0max}}\cos\alpha_{r}\cdot \mathrm{d}z/(z_{0max}-z_{0min})$. Here, it is assumed that the electrons, with initial position from $z_{0min}$ to $z_{0max}$, can be injected into the bubble. The $2$D theoretical model is in agreement with the $2$D PIC simulations, as presented at Fig. \ref{fig5}(d).

\begin{figure}[t]
\includegraphics[width=0.49\textwidth]{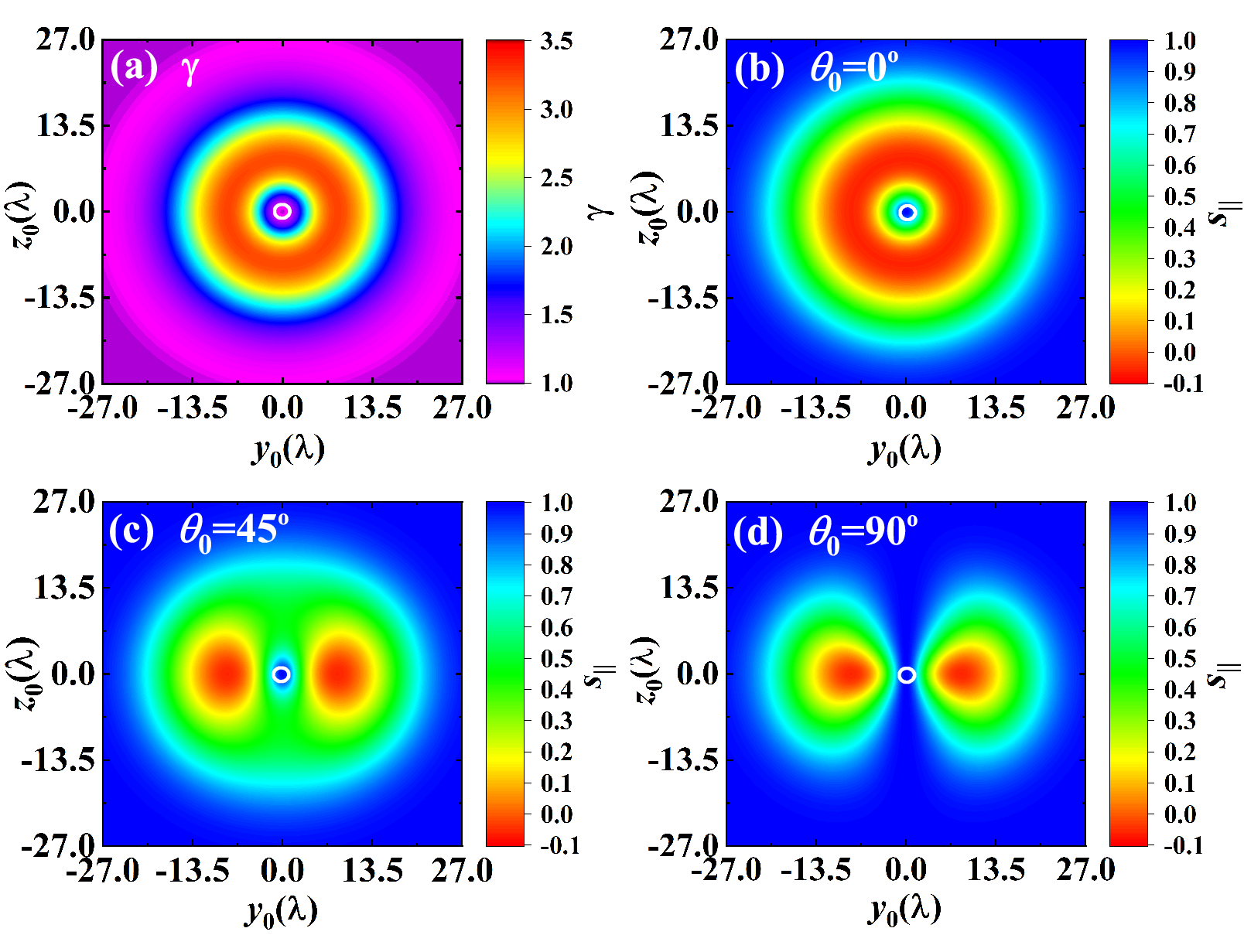}
\caption{(a) The distribution of final energy $\gamma$ for the electrons with their initial position ($y_{0},z_{0}$) after interacting with the laser in vacuum, calculated by the test particle dynamics. (b)-(c) The distributions of $s_{\parallel}$ for the electrons with their initial position ($y_{0},z_{0}$) under the different initial polarization directions $\theta_{0}=0^{\circ}$(b), $45^{\circ}$(c) and $90^{\circ}$(d). The laser parameters are $a_{0}=6$, $w_{0}=20\lambda$, and $\tau=17$fs. The initial position within $r_{0}=\pm0.8\lambda$ for electrons are marked by the white circles.}\label{fig6}
\end{figure}

\subsection{Longitudinal self-injection mechanism}

The self-injection mechanism can be switched between the transverse and longitudinal schemes by changing the plasma profile and/or the laser parameters. For longitudinal injection, the electron spin is mainly affected by the laser field, based on the work of Ref. \cite{L.R.Yin2024}. Considering that the laser evolution in the plasma is a $3$D phenomenon, it is difficult to give out the related expression for laser field. In addition, the laser profile does not vary significantly during the interaction with the injected electrons. Then, it is reasonable to consider the effects of the electron-laser interaction in vacuum. The electric field in $y$-direction of laser can be written as the Eq. (\ref{eq7}). The other electric and magnetic field components can be obtained using $E_{x}=(i/k_{0})\left(\partial E_{y}/\partial y\right)$, $\boldsymbol{B}=-(i/\omega_{0})\bigtriangledown\times \boldsymbol{E}$ based on the paraxial approximation \cite{P.X.Wang2002,Q.Kong}. Similar to the case of bubble fields, the electrons are at rest in front of the laser field with a position $(x_{0},y_{0},z_{0})$ and the electron spin $\boldsymbol{s}$ is located in the $xy$ plane initially, \emph{i.e.,} $\beta_{0}=0^{\circ}$. The laser parameters are $a_0=6$, $w_0=20\lambda$ and $\tau=17$ fs, respectively.

After interacted with laser, the $s_{\parallel}$ distributions of electrons at finial time $t=300$fs in dependence on their initial positions are shown in Figs. \ref{fig6}(c)-\ref{fig6}(d) for initial spin directions $\theta_{0}=0^{\circ}$, $45^{\circ}$, and $90^{\circ}$, respectively. The spins of electrons, near the laser axis initially, remain unchanged basically after undergoing the laser pulse in any case, as indicated by the white circle. During the longitudinal injection process, the trajectories of the accelerated electrons are close to the laser axis, which results in the fact that the net depolarization effect is almost zero even though the laser field can cause some spin precession of the electrons. When the electrons reach the tail of the bubble, the influence of the bubble field on the spin precession can also be ignored, due to the transverse electromagnetic field and the transverse velocity of the electrons being very small based on the work of Ref. \cite{L.R.Yin2024}.

\begin{figure}[b]
	\includegraphics[width=0.4\textwidth]{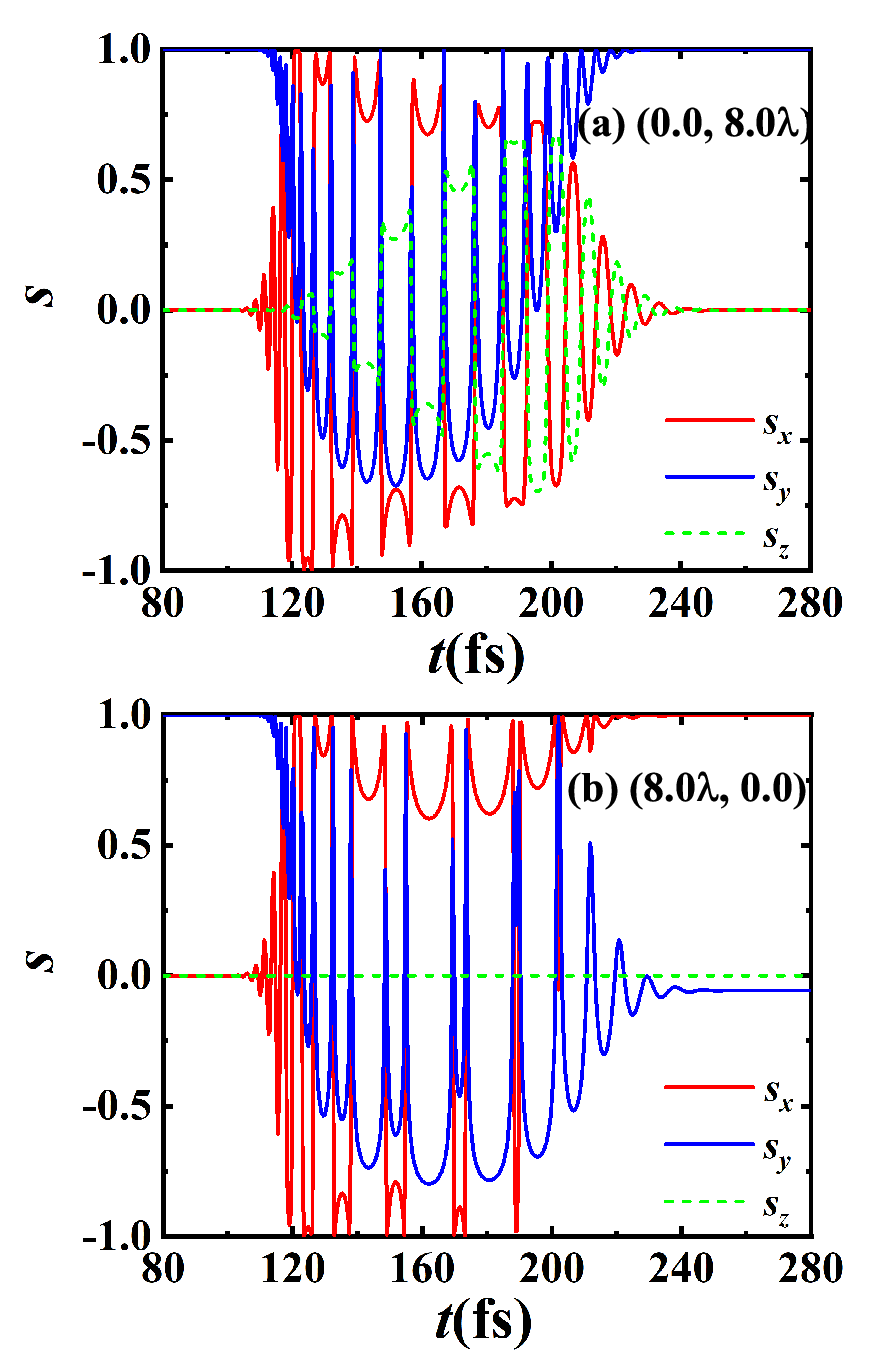}
	\caption{The history of electron spin at each direction for the typical electrons, where the initial positions ($y_{0},z_{0}$) of typical electrons are located at (0.0, 8.0$\lambda$) (a) and ($8.0\lambda$, 0.0) (b), respectively. The laser parameters and initial spin status are same as those in Fig. 6(d).}\label{fig7}
\end{figure}

With increasing the distance from the electron initial position to the laser axis, the situation alters obviously, as depicted in Fig. \ref{fig6}. The distribution of electron energy is axially symmetry, as revealed in Fig. \ref{fig6}(a), and the electron dynamics can be explained by the ponderomotive potential model. The transverse and longitudinal ponderomotive forces induced by the laser propagating along the $x$-axis can be derived as $F_{r}$ and $F_{x}$ \cite{D.Lin2013}. The transverse ponderomotive force is almost off-axis, the electron is always expelled transversely, while the longitudinal pondermotive force is basically positive and hence accelerates the electron longitudinally. Thus, the electrons have a transverse velocity $\widetilde{v_{r}}$ and a longitudinal velocity $\widetilde{v_{x}}$. The equivalent transversal ($\widetilde{E_{r}}$) and longitudinal electric fields ($\widetilde{E_{x}}$) are defined based on the ponderomotive model. More importantly, the electron velocity $\widetilde{v_{\varphi}}$ is zero, where the equivalent electric field $\widetilde{E_{\varphi}}$ is also zero.

\begin{figure}[t]
	\includegraphics[width=0.48\textwidth]{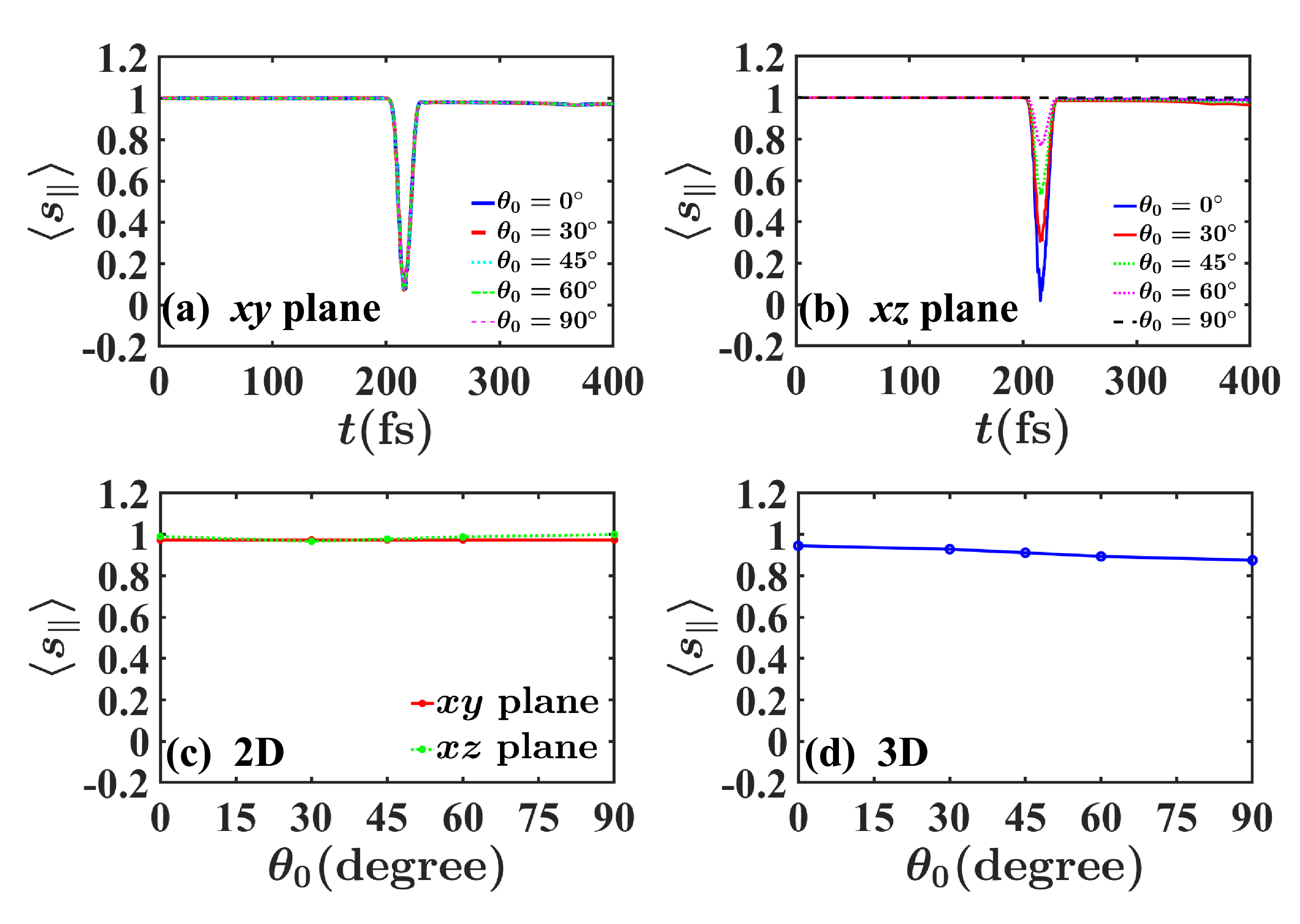}
	\caption{(a)-(b) The histories of polarization $\langle s_{\parallel}\rangle$ for the accelerating electrons under the longitudinal injection case with the different pre-polarization angle $\theta_{0}$ on $xy$ (a) and $xz$ planes (b) in $2$D PIC simulation. The relation between the initial spin direction $\theta_{0}$ and $\langle s_{\parallel}\rangle$ of the accelerating electrons in $2$D PIC simulation and the corresponding results of $3$D PIC simulation are plotted in (c) and (d), respectively. Here, the laser parameters are $a_{0}=6$, $w_{0}=20\lambda$ and $\tau=17$ fs and the plasma density is $n_{0}=0.040n_{c}$ and $L_{1}=45\lambda$.}\label{fig8}
\end{figure}	
	
During the interaction of the laser and the electrons, the electrons can gain net energy, and their spin will change. According to the TBMT equation, $\boldsymbol{\Omega} \propto \widetilde{\boldsymbol{v}}\times\widetilde{\boldsymbol{E}}$ and it can be derived as $\boldsymbol{\Omega}=\widetilde{\boldsymbol{\Omega_{\varphi}}}$. Therefore, the distribution of electron spin is axial symmetry at the case of $\theta_0=0^{\circ}$, as observed in Fig. \ref{fig6}(b). When changing the initial spin direction, the distribution of the finial electron spin is also altered, which is similar to the case of the bubble field, as illustrated in Figs. \ref{fig6}(c) and \ref{fig6}(d).

In order to confirm the theoretical analysis, the spin evolution of two electrons with typical initial positions is presented in Fig. \ref{fig7}. Here, the initial electron spin is aligned with the $y$-axis. The other parameters are same as those in Fig. \ref{fig6}(d). For the electrons initially located on the $z$-axis, the spin direction is always parallel to the precession frequency $\widetilde{\boldsymbol{\Omega_{\varphi}}}$. Then, $s_{y}$ quivers in the laser field, while the net depolarization is zero, as demonstrated in Fig. \ref{fig7}(a). Electrons initially located on the $y$-axis oscillates in the $xy$ plane while interacting with the laser field, and their spins are always perpendicular to the precession frequency $\widetilde{\boldsymbol{\Omega_{\varphi}}}$ and $s_{z}$ is always zero, as shown in Fig. \ref{fig7}(b).

Figure \ref{fig8} displays the related result of PIC simulation. Here, the initial laser amplitude is $a_{0}=6$, the laser waist is $w_{0}=20\lambda$, the laser pulse duration is $\tau=17$fs, the length of the up-ramp transition is $L_{1}=45\lambda$, the density of plateau is $n_{0}=0.04n_{c}$, and the other parameters are same as those in the case of transverse injection. Accelerated electrons with energy larger than $15$MeV at $500$fs are considered for the longitudinal injection process. As shown in Figs. \ref{fig8}(a) and \ref{fig8}(b), the $2$D simulation result confirm the theoretical analysis based on the pondermotive model. The effect of initial electron spin can be ignored at the final time. While the electrons interact with laser, the electron spin is affected by laser directly at around $200$fs. The electrons do not gain net energy after interacted with laser and the net effect of depolarization is nearly zero for the longitudinal injection scheme as presented in Figs. \ref{fig8}(c) and \ref{fig8}(d). Compared to the case of transverse injection, the longitudinal injection is more stable to obtain an electron beam with high polarization, especially independent on the direction of initial spin.

\section{Conclusion}
The effect of initial spin orientation on the polarization of accelerated electrons in the LWFA regime is investigated theoretically and numerically using the Thomas-Bargmann-Michel-Telegdi equation. It is found that the final spin polarization rate depends upon the self-injection processes, \emph{i.e.,} transverse self-injection and longitudinal self-injection, respectively. In the case of transverse injection, the electron spin is mainly affected by the bubble field and the direction of precession frequency is fixed in the vertical plane of the electron trajectory. The spins of several electrons do not move when their spins are initially aligned to the direction of precession frequency, and it causes the final polarization of the accelerated electron beam to depend on its initial spin status. However, in the case of longitudinal injection, the dynamics of electron spin is mainly determined by the laser field. Due to the net effect of laser field being negligible, the polarization of the accelerated electrons is independent on their initial spin status. The result of $3$D PIC simulation is consistent with the test particle simulation and theoretical analysis. This study indicates that the longitudinal self-injection mechanism is more stable to obtain a bunch of electrons with high polarization, compared to the case of transverse self-injection. Furthermore, the study of bubble and laser fields about electron spin is also useful to investigate the polarization dynamics in other acceleration schemes.

\section{ACKNOWLEDGEMENTS}
This work was supported by the National Natural Science Foundation of China (No.11804348, No.11775056, No.11975154, No.12225505 and No.12405281), and the Science Challenge Project (No.TZ2018005). X. F. Li was also supported by Shanghai Pujiang Program (Grant No. 23PJ1414600). The work of M. B. was carried out in the framework of the J\"ulich Short-Pulse Particle and Radiation Center \cite{buscher2020jusparc} and was supported by the Accelerator Technology Helmholtz Infrastructure consortium ATHENA.


\end{document}